\documentclass[conference]{IEEEtran}

\usepackage[cmex10]{amsmath}
\usepackage{amsfonts,amssymb,amsbsy}
\usepackage{graphicx,stfloats,comment}

\newcommand{\set}[1]{\mathcal{#1}} 

\newcommand{\ul}[1]{\underline{#1}}
\newcommand{\E}[1]{{\rm E}\left[{#1}\right]}
\newcommand{\Tr}[1]{{\rm Tr}\left({#1}\right)}
\newcommand{\sinc}{{\rm sinc}}

\begin{document}

\author{
\IEEEauthorblockN{Gerhard Kramer$^*$, Mansoor I. Yousefi$^*$, and Frank R. Kschischang$^\dag$}
\IEEEauthorblockA{$^*$Institute for Communications Engineering, Technische Universit\"{a}t M\"{u}nchen, Germany \\
$^\dag$Edward S. Rogers Sr. Dept. of Electrical \& Computer Engineering, University of Toronto, Canada}
}

\title{Upper Bound on the Capacity of a Cascade \\ of Nonlinear and Noisy Channels}

\maketitle

\begin{abstract}
An upper bound on the capacity of a cascade of nonlinear and noisy channels is presented. The cascade mimics the split-step Fourier method for computing waveform propagation governed by the stochastic generalized nonlinear Schr\"{o}dinger equation. It is shown that the spectral efficiency of the cascade is at most log(1+SNR), where SNR is the receiver signal-to-noise ratio. The results may be applied to optical fiber channels. However, the definition of bandwidth is subtle and leaves open interpretations of the bound. Some of these interpretations are discussed.
\end{abstract}


\section{Introduction}
\label{sec:intro}
The capacity of the optical fiber channel seems difficult to compute or even bound.
Perhaps the best-known capacity {\em lower} bound for optical fiber {\em networks} is given
in~\cite{Essiambre-JLT10}. Many follow-up papers have suggested modifications
of this bound, e.g. see~\cite{Bosco-OE11,Bosco-OE12,Poggiolini-JLT12,Mecozzi-JLT12,Secondini-JLT13,Dar-JLT13,Agrell-JLT14,Yousefi-IT14} and references therein. 
The purpose of this paper is to develop a simple capacity {\em upper} bound for a
class of channels that is sometimes used to simulate signal propagation.
As far as we know, this is the first capacity upper bound for the optical fiber channel.
The bound is based on two basic tools:\
maximum entropy under a correlation constraint and the entropy power inequality (EPI).
The main insight is that the non-linearity that is commonly used to model optical fiber
propagation does not change the differential entropy of a signal.

The capacity bound can be converted to a spectral efficiency bound by normalizing
by an appropriate bandwidth. We caution, however, that it is not clear what the
``right" choice for normalization should be. We therefore consider several options
that may or may not be satisfactory for the engineering problem. We discuss
extensions to the basic model that might help to clarify the issue.

This paper is organized as follows. In Sec.~\ref{sec:prelim} we review basic results on 
complex random variables and their entropy. In Sec.~\ref{sec:signal} we consider continuous and
discrete signal propagation models for optical fiber. In Sec.~\ref{sec:capacity} we
develop an upper bound on capacity for the discretized model. In Sec.~\ref{sec:discussion} we
discuss subtleties concerning how to normalize capacity to compute a bound on spectral
efficiency. We further outline some extensions. Sec.~\ref{sec:conclusions}
concludes the paper. 

We remark that the methods presented here were recently adapted to an
optical fiber model that is based on Hamiltonian energy-preserving
dynamical systems~\cite{Yousefi-CWIT15}.

\section{Preliminaries}
\label{sec:prelim}
\subsection{Proper Complex Random Variables}
Let $j=\sqrt{-1}$ and consider a complex random column vector
$\ul{X}=\ul{X}_c+j\ul{X}_s$ where $\ul{X}_c$ and $\ul{X}_s$
are real random column vectors. The covariance and pseudo-covariance
matrices of $\ul{X}$ are defined as the respective
\begin{align}
   & {\bf Q}_{\ul{X}} = \E{(\ul{X}-\E{\ul{X}})(\ul{X}-\E{\ul{X}})^\dag} \\
   & {\bf \tilde{Q}}_{\ul{X}} = \E{(\ul{X}-\E{\ul{X}})(\ul{X}-\E{\ul{X}})^T}
\end{align}
where $\ul{X}^T$ and $\ul{X}^\dag$ are the respective transpose and
complex-conjugate transpose of $\ul{X}$. The complex vector $\ul{X}$
is said to be {\it proper} if ${\bf \tilde{Q}}_{\ul{X}}={\bf 0}$
(see \cite{Neeser93}). It is known that a linear or affine transformation
of a proper complex $\ul{X}$ is also proper~\cite[Lemma~3]{Neeser93}.

\subsection{Differential Entropy}
Let $h(\ul{X})=h(\ul{X}_c\,\ul{X}_s)$ be the differential entropy of $\ul{X}$.
Two basic properties of $h(\cdot)$ are the translating and scaling properties:\
for a complex vector $\ul{v}$ and a complex square matrix ${\bf M}$ we have
\begin{align}
    & h(\ul{X}+\ul{v}) = h(\ul{X}) \label{eq:translation} \\
    & h({\bf M} \ul{X}) = h(\ul{X}) + 2 \log |\det(\bf M)| \label{eq:scaling}
\end{align}
where $\det({\bf M})$ is the determinant of ${\bf M}$ and $|X|$ is the
absolute value of $X$.  For instance, if ${\bf M}$ is unitary
(${\bf M}^{-1}={\bf M}^\dag$) then we have $|\det({\bf M})|=1$ and
$h({\bf M} \ul{X})=h(\ul{X})$. We remark that real random vectors and
real matrices do not have the factor 2 in \eqref{eq:scaling},
see~\cite[Eq.~(13)]{Neeser93}.

\subsection{Discrete Fourier Tranform}
The discrete Fourier transform (DFT) of a $L\times1$ vector
$\ul{a}$ is $\ul{A}={\bf F}\,\ul{a}$ where ${\bf F}$ is the $L \times L$
discrete Fourier transform (DFT) matrix with entries
\begin{align*}
   \frac{1}{\sqrt{L}} \, e^{-j 2 \pi \ell m / L }, \quad 0 \le \ell,m \le L-1 .
\end{align*}
Observe that ${\bf F}$ is unitary (${\bf F}^{-1}={\bf F}^\dag$). The inverse
Fourier transform (IDFT) of $\ul{A}$ is $\ul{a}={\bf F}^\dag \, \ul{A}$.

\subsection{Maximum Entropy}
A useful property of $h(\cdot)$ is a maximum entropy result proved in
~\cite[Thm.~2]{Neeser93}: for a $L\times1$ complex vector $\ul{X}$ with nonsingular
correlation matrix ${\bf R}(\ul{X}) := \E{\ul{X} \, \ul{X}^\dag}$ we have
\begin{align} \label{eq:max-ent}
   h(\ul{X}) \le \log [(\pi e)^L \det({\bf R}(\ul{X}))]
\end{align}
with equality if and only if $\ul{X}$ is proper complex,
Gaussian, and zero mean.

\subsection{Entropy Power}
The entropy power of a real, random vector $\ul{X}$ of length $L$
is defined as $V(\ul{X})=e^{2h(\ul{X})/L}/(2\pi e)$. But a complex vector $\ul{X}$
of length $L$ can be considered to be a real vector of length $2L$, so
when dealing with complex vectors we instead use the definition
$V(\ul{X})=e^{h(\ul{X})/L}/(\pi e)$. 
So consider two independent, complex, random vectors $\ul{X}$ and
$\ul{Y}$ of length $L$. The EPI states that~\cite[Sec. 17.8]{Cover06}
\begin{align} \label{eq:epi}
   V\left(\ul{X} + \ul{Y}\right) \ge V(\ul{X}) + V(\ul{Y}).
\end{align}

\section{Signal Propagation}
\label{sec:signal}
\subsection{Continuous Space-Time Equations}
Suppose the signal $a(z,t)$ represents the optical field at location $z$ and time $t$.
The location $z=0$ usually represents the launch position, i.e., the launch
signal is $a(0,t)$. We take the receive signal to be at position $z^*$,
i.e., the receive signal is $a(z^*,t)$.
For ideal distributed Raman amplification (see~\cite[Sec.~IX.B]{Essiambre-JLT10})
the evolution of $a(z,t)$ is given by the generalized nonlinear Schr\"{o}dinger equation
(see~\cite[eq.~(70)]{Essiambre-JLT10})
\begin{align}
   \frac{\partial a}{\partial z} + j \frac{\beta_2}{2} \frac{\partial^2 a}{\partial t^2} - j \gamma |a|^2 a
   & =  n
   \label{eq:NLSE}
\end{align}
where $j=\sqrt{-1}$ and $n$ is a Gaussian noise process that is spatially white and
bandlimited to $B_n$ Hertz. In other words, the noise spatial and temporal autocorrelation
function is (see~\cite[eq.~(53)]{Essiambre-JLT10})
\begin{align}
   \E{n(z,t)n(z',t')^*} = \frac{N_{\rm ASE}}{z^*} \, \delta(z-z') \, B_n \sinc\left(B_n (t-t')\right)
   \label{eq:noise-a}
\end{align}
where $\delta(x)$ is the Dirac-delta generalized function and $\sinc(x)=\sin(\pi x)/(\pi x)$.

\subsection{Discrete Space-Time Equations}
The evolution of $a(z,t)$ in \eqref{eq:NLSE} is often computed by
using the split-step Fourier method. This method discretizes both
space and time, i.e., $z$ takes on values in the set 
$\set{Z}=\{z_0,z_1,\ldots,z_K\}$ and $t$ takes on values in the set
$\set{T}=\{t_0,t_1,\ldots,t_{L-1}\}$. Usually the space and time values
are chosen to be uniformly-spaced, i.e., $z_k=\Delta_z k$ and
$t_\ell = \Delta_t \ell$ for some constants $\Delta_z$ and $\Delta_t$
and integers $k=0,1,\ldots,K$, $\ell=0,1,\ldots,L-1$.
We will use this simplification below, although
a more general approach with non-uniform spacing is possible.
We write $\ul{a}(z_k)$ for the $L \times 1$ vector of
sample values $a(z_k,t_{\ell})$, $\ell=0,1,\ldots,L-1$, at position $z_k$.

The evolution of $\ul{a}(z)$ from position $z=0$ to position
$z=z^*$ is performed by recursively computing $K$ ``small'' steps
from position $z_k$ to $z_{k+1}$ for $k=0,1,\ldots,K-1$.
More precisely, the signal evolution from position $z_k$ to position $z_{k+1}$ is
computed by ``splitting" the linear and nonlinear steps.
\begin{enumerate}
\item Nonlinear step. Compute the effect of nonlinearity via
\begin{align}
   \ul{a}_N(z_{k+1}) = {\bf D}_N \, \ul{a}(z_k)
\end{align}
where ${\bf D}_N$ is a diagonal matrix with entries
\begin{align}
   e^{j \gamma \, |a(z_k,t_\ell)|^2 \, \Delta_z}, \quad \ell=0,1,\ldots,L-1
\end{align}
and where $a(z_k,t_\ell)$ is the $(\ell+1)$ entry of $\ul{a}(z_k)$.
\item Linear step. Use the DFT to compute
$\ul{A}_N(z_{k+1}) =F \, \ul{a}_N(z_{k+1})$. Next compute the effect of
dispersion via
\begin{align}
   \ul{A}_L(z_{k+1}) = {\bf D}_L \, \ul{A}_N(z_{k+1})
\end{align}
where ${\bf D}_L$ is a diagonal matrix with entries
\begin{align}
   & e^{-j (\beta_2/2) \, \ell^2/(L\Delta_t)^2 \, \Delta_z }, \; \ell=0,1,\ldots,L/2-1 \\
   & e^{-j (\beta_2/2) \, (L-\ell)^2/(L\Delta_t)^2 \, \Delta_z }, \; \ell=L/2,\ldots,L-1
\end{align}
where we assumed that $L$ is even.
Finally, use the IDFT to compute $\ul{a}_L(z_{k+1}) ={\bf F}^{\dag} \, \ul{A}_L(z_{k+1})$.
Summarizing, the linear step has input $\ul{a}_N(z_{k+1})$ and output
\begin{align}
   \ul{a}_L(z_{k+1}) ={\bf F}^\dag \, {\bf D}_L \, {\bf F} \, \ul{a}_N(z_{k+1}).
\end{align}
\item Noise step. Add noise whose variance is proportional
to the space step $\Delta_z$, the time step $\Delta_t$, and the noise bandwidth $B_n$.
We assume that the simulation bandwidth $B=1/\Delta_t$ satisfies $B \ll B_n$.
We compute the effect of noise via
\begin{align} \label{eq:noise-step}
   \ul{a}(z_{k+1}) = \ul{a}_L(z_{k+1}) + \ul{n}(z_{k+1})
\end{align}
where the entries of the $L\times1$ column vector $\ul{n}(z_{k+1})$ are drawn
independently from a proper complex Gaussian distribution with
variance $(N_{\rm ASE} B_n/z^*) \Delta_z \Delta_t$.
\end{enumerate}
In summary, one step in space requires computing
\begin{align}
   \ul{a}(z_{k+1}) = {\bf F}^\dag \, {\bf D}_L \, {\bf F} \, {\bf D}_N \, \ul{a}(z_k) + \ul{n}(z_{k+1}).
   \label{eq:stepk}
\end{align}
Although this equation looks linear, the nonlinearity arises because ${\bf D}_N$
depends on the $|a(z_k,t_\ell)|^2$ for all $\ell$.

We remark that several split-step methods can be used, e.g., one can use two
(fine) linear steps and one nonlinear step. The motivation for doing this is to improve
numerical accuracy and/or speed up simulations. The choice of method does
not affect the results below.

\section{Capacity Bound}
\label{sec:capacity}

We develop an upper bound on the mutual information $I(\ul{a}(0) ; \ul{a}(z^*) )$
between the channel input and output signals. The bound uses two basic ideas.
\begin{enumerate}
\item Compute the energy of the output signal and apply the maximum entropy bound~\eqref{eq:max-ent}.
\item Show that the nonlinear step does not change entropy and apply the EPI \eqref{eq:epi}.
\end{enumerate}

\subsection{Output Energy}
\label{subsec:covariance}
Consider the space step from position $z_k$ to position $z_{k+1}$. The correlation matrix
${\bf R}(\ul{a}_N(z_{k+1}))$ of $\ul{a}_N(z_{k+1})$ has as $(\ell,m)$
entry the value
\begin{align}
    & \E{ a_N(z_{k+1},t_\ell) a_N(z_{k+1},t_m)^* } = \nonumber \\
    & \E{ a(z_k,t_\ell) a(z_k,t_m)^* e^{j 2 \pi \gamma (|a(z_k,t_\ell)|^2-|a(z_k,t_m)|^2) \Delta_z } }
\end{align}
Observe that the $m=\ell$ entries do not change, i.e., the diagonal of ${\bf R}(\ul{a}_N(z_{k+1}))$
is the same as the diagonal of ${\bf R}(\ul{a}(z_{k}))$. We thus have $\Tr{{\bf R}(\ul{a}_N(z_{k+1}))}=\Tr{{\bf R}(\ul{a}(z_{k}))}$,
where $\Tr{\bf M}$ is the trace of the square matrix $\bf M$.
Next, we compute
\begin{align}
    {\bf R}(\ul{a}_L(z_{k+1})) = {\bf F}^\dag {\bf D}_L {\bf F} \, {\bf R}(\ul{a}_N(z_{k+1})) \, {\bf F}^\dag {\bf D}_L^\dag {\bf F}.
\end{align}
We thus have $\Tr{{\bf R}(\ul{a}_L(z_{k+1}))}=\Tr{{\bf R}(\ul{a}_N(z_{k+1}))}$ by repeatedly using $\Tr{{\bf A}{\bf B}}=\Tr{{\bf B}{\bf A}}$.
Finally, we have
\begin{align}
    {\bf R}(\ul{a}(z_{k+1})) = {\bf R}(\ul{a}_L(z_{k+1})) + \frac{N_{\rm ASE} B_n}{z^*} \Delta_z \Delta_t \: {\bf I}
\end{align}
where ${\bf I}$ is the $L \times L$ identity matrix.

Combining the above results, we have
\begin{align}
  & \Tr{{\bf R}(\ul{a}(z_{K}))} = \Tr{ {\bf R}(\ul{a}(z_{0})) } + L N_{\rm ASE} B_n \Delta_t \nonumber \\
  & = E_0 +  N_{\rm ASE} B_n T
\end{align}
where $E_0 = \Tr{ {\bf R}(\ul{a}(z_{0}))}$ is the input signal energy and $T=L\Delta_t$ is the total time.
We further have
\begin{align}
    & \log \det {\bf R}(\ul{a}(z_K))
       \overset{(a)}{\le} \log \left( \prod_{i=1}^L R_{i,i}(\ul{a}(z_K)) \right) \nonumber \\
    & = \sum_{i=1}^L \log R_{i,i}(\ul{a}(z_K)) \nonumber \\
    & \overset{(b)}{\le} L \log\left( \Tr{{\bf R}(\ul{a}(z_K))}/L \right) \nonumber \\
    & = L \log\left( (E_0 + N_{\rm ASE}B_nT)/L \right)
\end{align}
where $(a)$ follows by defining $R_{i,i}(\ul{a}(z_K))=\E{|a(z_K,t_{i-1})|^2}$ as the $(i,i)$ entry of ${\bf R}(\ul{a}(z_K))$ and applying
Hadamard's inequality~\cite[Sec. 17.9]{Cover06}, and $(b)$ follows by Jensen's inequality.
Using~\eqref{eq:max-ent}, we thus have the entropy upper bound
\begin{align} \label{eq:upper}
    h(\ul{a}(z^*)) \le L \log\left( \pi e \left( E_0 + N_{\rm ASE} B_n T \right)/L \right) .
\end{align}

\subsection{Entropy Preservation}
\label{subsec:entropies}
The linear step preserves entropy because ${\bf D}_L$ is a unitary matrix.
For the nonlinearity, observe that every entry of $\ul{a}_N(z_{k+1})$ has the form $|a| e^{j \arg(a) + j f(|a|)}$ where
$\arg(a)$ is the phase of $a$ and $f$ is a smooth function. We compute
\begin{align}
   & h\left( |a| e^{j \arg(a) + j f(|a|)} \right) \nonumber \\
   & \overset{(a)}{=} h\left( |a|, \arg(a) + f(|a|) \text{ mod } 2 \pi  \right) + \E{\log |a|} \nonumber \\
   & \overset{(b)}{=} h\left( |a| \right) + h\left(\arg(a) + f(|a|) \text{ mod } 2 \pi  \big| |a| \right) + \E{\log |a|} \nonumber \\
   & \overset{(c)}{=} h\left( |a| \right) + h\left(\arg(a) \text{ mod } 2 \pi  \big| |a| \right) + \E{\log |a|} \nonumber \\
   & = h\left( |a| e^{j \arg(a)} \right) \nonumber \\
   & = h\left( a \right) \label{eq:entropy-equal}
\end{align}
where $(a)$ follows by \cite[eq: (318)]{Lapidoth:03}, $(b)$ follows by the chain rule for entropy,
and $(c)$ follows by \eqref{eq:translation}.
The above steps remain valid with conditioning. We thus have
\begin{align}
  & h(\ul{a}_N(z_{k+1}) | \ul{a}(0)) \nonumber \\
  & = \sum_{\ell=0}^{L-1} h(a_N(z_{k+1},t_{\ell}) | \ul{a}(0), a_N(z_{k+1},t_{0}), \ldots, a_N(z_{k+1},t_{\ell-1})    ) \nonumber \\
  & \overset{(a)}{=} \sum_{\ell=0}^{L-1} h(a(z_k,t_{\ell}) | \ul{a}(0), a(z_{k},t_{0}), \ldots, a(z_{k},t_{\ell-1})    ) \nonumber \\
  & = h(\ul{a}(z_{k}) | \ul{a}(0)) 
\end{align}
where $(a)$ follows because there is an invertible transformation from $a_N(z_{k+1},t_{\ell})$
to $a(z_{k},t_{\ell})$ for all $\ell$, and hence we can exchange these values when conditioning.

The above results imply that
\begin{align}
  & V(\ul{a}(z_{k+1}) | \ul{a}(0)) \nonumber \\
  & = V\left( \ul{a}_L(z_{k+1}) + \ul{n}(z_{k+1}) \left| \: \ul{a}(0) \right. \right) \nonumber \\
  & \overset{(a)}{\ge} V(\ul{a}_L(z_{k+1}) | \ul{a}(0)) + V(\ul{n}(z_{k+1}) | \ul{a}(0)) \nonumber \\
  & =  V(\ul{a}(z_{k}) | \ul{a}(0)) + (N_{\rm ASE} B_n / z^*) \Delta_z \Delta_t
\end{align}
where $(a)$ follows by \eqref{eq:epi}. By induction, we thus have
$V(\ul{a}(z_K) | \ul{a}(0)) \ge N_{\rm ASE}B_n \Delta_t$
and therefore
\begin{align} \label{eq:lower}
    h(\ul{a}(z^*) | \ul{a}(0)) \ge L \log\left( \pi e N_{\rm ASE} B_n T/L \right) .
\end{align}
Combining the results \eqref{eq:upper} and \eqref{eq:lower}, we have
\begin{align} \label{eq:C-bound}
    I(\ul{a}(0) ; \ul{a}(z^*) ) \le L \log\left( 1 + \frac{E_0}{N_{\rm ASE} B_n T} \right) .
\end{align}

\section{Discussion}
\label{sec:discussion}
%
\subsection{Spectral Efficiency}
\label{subsec:SE}
The bound \eqref{eq:C-bound} normalized by the time $T=L/B$ states that the
capacity is upper bounded as
\begin{align} \label{eq:C-bound2}
   \widetilde{C} & \le B \log\left( 1 + {\rm SNR}\right) \quad \text{bits/s}
\end{align}
where ${\rm SNR}=E_0/(N_{\rm ASE} B_n T)$ is the signal-to-noise ratio.

To bound the spectral efficiency, we must normalize by the bandwidth.
How to define bandwidth precisely is open to interpretation, but suppose for
now that $\ul{a}(z_k)$ has bandwidth $W(z_k)$. For example, $W(z_k)$ could
be the smallest bandwidth for which the simulations do not substantially
reduce the mutual information $I(\ul{a}(z_0) ; \ul{a}(z_k) )$. Another approach
is to choose $W(z_k)$ large enough to ensure that the ``out-of-band interference"
is ``sufficiently weak", since interference plays a major role when sharing spectrum.

We use the following approach.
The maximal signal bandwidth is $W=\max_{0\le k\le K} W(z_k)$ and we define
the spectral efficiency to be
\begin{align}
   C & \le \frac{B}{W} \log\left( 1 + {\rm SNR}\right) \quad \text{bits/s/Hz}.
\end{align}
The best bound follows by choosing $B=W$. We feel that this is the ``right"
approach because the frequency band with bandwidth $W$ carries (almost) all the mutual information,
and because one usually assumes that only the signals of interest are present inside this band.
The spectral efficiency is thus bounded by $\log(1+{\rm SNR})$.

It remains to be seen whether our approach will be generally accepted. In any case,
the upper bound \eqref{eq:C-bound2} on {\em capacity} remains valid.
As far as we know, this is the first such bound for optical fiber channels.

\subsection{Extensions}
\label{subsec:extend}
There are several possible extensions of the results. First, observe that for the
nonlinear step the phase can be any function of the amplitude
without changing \eqref{eq:entropy-equal}. One will usually choose a smooth function
so that the discrete version of the problem matches the continuous version.

Second, observe that for the linear step one may choose any unitary transform, i.e.,
any all-pass filter. For instance, the bound derived above remains valid for third-order dispersion.

Third, one may wish to study a model where Raman amplification takes place
in the C-band (see~\cite[Fig.~21]{Essiambre-JLT10}) but
advanced receivers collect and process signals both inside and outside this band.
Outside the C-band, the signal will experience loss of at least 0.2 dB/km which is
at least 20 dB for 100 km. However, the noise outside the C-band is weak so that
out-of-band processing could be interesting. In this case, the noise step \eqref{eq:noise-step}
should be modified to include a frequency-dependent loss and noise variance. This model
is realistic for heterogeneous systems where amplification is
frequency dependent, and where receivers process outside of the bandwidth of their
signals of interest. We hope that more detailed models such as this one could help
to clarify what a proper definition of bandwidth might be.

Finally, the bound extends to problems with polarization, core, and
mode multiplexing, as long as the linear and non-linear steps preserve energy and entropy,
and the noise is additive, independent, and Gaussian.

\section{Conclusions}
\label{sec:conclusions}
The spectral efficiency of a cascade of nonlinear and noisy channels was shown to be bounded
by $\log(1+{\rm SNR})$. The definition of spectral efficiency is subtle, however, because
the notion of bandwidth is subtle, e.g., see~\cite{Slepian76}. In fact, a recent
paper~\cite{Terekhov14} states that the spectral efficiency of the optical fiber
channel can be larger than $\log(1+{\rm SNR})$ when one normalizes by the
input bandwidth $W(z_0)$ (see the text after (16) in~\cite{Terekhov14}).
Of course, $W(z_0)$ is generally smaller than the maximal bandwidth $W$.
This observation is important because, among other considerations, the
maximal bandwidth defines the required spectral spacing when using
wavelength-division multiplexing (WDM).

\section*{Acknowledgment}
\label{sec:ack}
G. Kramer and M.~I. Yousefi were supported by an Alexander von Humboldt Professorship endowed
by the German Federal Ministry of Education and Research. M.~I. Yousefi and F.~R. Kschischang
were supported by the Technische Universit\"{a}t M\"{u}nchen Institute for Advanced Study (TUM-IAS)
in the framework of a Hans Fischer Senior Fellowship.

\bibliographystyle{IEEE} 

\begin{thebibliography}{10}

\bibitem{Essiambre-JLT10}
R.-J. Essiambre, G.~Kramer, P.~J. Winzer, G.~J. Foschini, and B.~Goebel,
\newblock ``Capacity limits of optical fiber networks,''
\newblock {\em IEEE/OSA J.\ Lightwave Techn.}, vol. 28, no. 4, pp. 662--701,
  Feb. 2010.

\bibitem{Bosco-OE11}
G.~Bosco, P.~Poggiolini, A.~Carena, V.~Curri, and F.~Forghieri,
\newblock ``Analytical results on channel capacity in uncompensated optical
  links with coherent detection,''
\newblock {\em Optics Express}, vol. 19, no. 26, pp. B440--B451, Dec. 2011.

\bibitem{Bosco-OE12}
G.~Bosco, P.~Poggiolini, A.~Carena, V.~Curri, and F.~Forghieri,
\newblock ``Analytical results on channel capacity in uncompensated optical
  links with coherent detection: erratum,''
\newblock {\em Optics Express}, vol. 20, no. 17, pp. 19610--19611, Aug. 2012.

\bibitem{Poggiolini-JLT12}
P.~Poggiolini,
\newblock ``The {GN} model of non-linear propagation in uncompensated coherent
  optical systems,''
\newblock {\em IEEE/OSA J.\ Lightwave Techn.}, vol. 30, no. 24, pp. 3857--3879,
  Dec. 2012.

\bibitem{Mecozzi-JLT12}
A.~Mecozzi and R.-J. Essiambre,
\newblock ``Nonlinear {S}hannon limit in pseudolinear systems,''
\newblock {\em IEEE/OSA J.\ Lightwave Techn.}, vol. 30, no. 12, pp. 2011--2024,
  June 2012.

\bibitem{Secondini-JLT13}
M.~Secondini, E.~Forestieri, and G.~Prati,
\newblock ``Achievable information rate in nonlinear {WDM} fiber-optic systems
  with arbitrary modulation formats and dispersion maps,''
\newblock {\em IEEE/OSA J.\ Lightwave Techn.}, vol. 31, no. 23, pp. 3839--3852,
  Dec. 2013.

\bibitem{Dar-JLT13}
R.~Dar, M.~Shtaif, and M.~Feder,
\newblock ``New bounds on the capacity of fiber-optics communications,''
\newblock {\em ArXiv e-prints}, 2013,
\newblock Online at: http://arxiv.org/abs/1305.1762.

\bibitem{Agrell-JLT14}
E.~Agrell, A.~Alvarado, G.~Durisi, and M.~Karlsson,
\newblock ``Capacity of a nonlinear optical channel with finite memory,''
\newblock {\em IEEE/OSA J.\ Lightwave Techn.}, vol. 32, no. 16, pp. 2862--2876,
  Aug. 2014.

\bibitem{Yousefi-IT14}
M.~I. Yousefi and F.~R. Kschischang,
\newblock ``Information transmission using the nonlinear {F}ourier transform:
  {P}arts {I-III},''
\newblock {\em IEEE Trans.\ Inf.\ Theory}, vol. 60, no. 7, pp. 4312--4369, July
  2014.

\bibitem{Yousefi-CWIT15}
M.~I. Yousefi, G.~Kramer, and F.~R. Kschischang,
\newblock ``An upper bound on the capacity of the single-user nonlinear
  {S}chr{\"{o}}dinger channel,''
\newblock in {\em Can. Workshop Inf. Theory}, St. John's, NL, Canada, July 6-9
  2015,
\newblock Online at: http://arxiv.org/abs/1502.06455v1.

\bibitem{Neeser93}
F.D. Neeser and J.L. Massey,
\newblock ``Proper complex random processes with applications to information
  theory,''
\newblock {\em IEEE Trans.\ Inf.\ Theory}, vol. 39, pp. 1293--1302, July 1993.

\bibitem{Cover06}
T.~M. Cover and J.~A. Thomas,
\newblock {\em Elements of Information Theory},
\newblock John Wiley \& Sons, New York, 2nd edition, 2006.

\bibitem{Lapidoth:03}
A.~Lapidoth and S.~M. Moser,
\newblock ``Capacity bounds via duality with applications to multiple-antenna
  systems on flat fading channels,''
\newblock {\em IEEE Trans.\ Inf.\ Theory}, vol. 49, no. 10, pp. 2426--2467,
  Oct. 2003.

\bibitem{Slepian76}
D.~Slepian,
\newblock ``On bandwidth,''
\newblock {\em Proc. IEEE}, vol. 64, pp. 292--300, Mar. 1976.

\bibitem{Terekhov14}
I.~S. Terekhov, A.~V. Reznichenko, and S.~K. Turitsyn,
\newblock ``Mutual information in nonlinear communication channel: analytical
  results in large {SNR} limit,''
\newblock {\em ArXiv e-prints}, 2013,
\newblock Online at: http://arxiv.org/abs/1411.7477.

\end{thebibliography}

\end{document}